\documentclass[superscriptaddress]{revtex4}
\usepackage{graphicx}
\usepackage{amsmath}
\usepackage{color}
\usepackage{amssymb}

\begin{document}
\title{Theory of Spoof Magnetic Localized Surface Plasmons Beyond Effective Medium Approximations}

\author{Carlo Rizza}
\affiliation{Department of Physical and Chemical Sciences, Via Vetoio 1, University of L'Aquila, I-67100 L'Aquila, Italy}
\affiliation{Institute for Superconductors, Oxides and Other Innovative Materials and Devices, National Research Council, Via Vetoio 1, I-67100 L'Aquila, Italy}
\author{Angelo Galante}
\affiliation{Department of Life, Health and Environmental Sciences, University of L'Aquila, Via Vetoio 1, I-67100 L'Aquila, Italy} 
\affiliation{Institute for Superconductors, Oxides and Other Innovative Materials and Devices, National Research Council, Via Vetoio 1, I-67100 L'Aquila, Italy}
\affiliation{National Institute for Nuclear Physics
(INFN), Gran Sasso National Laboratory, I-67100,  L'Aquila, Italy}
\author{Elia Palange}
\affiliation{Department of Industrial and Information Engineering and Economics, Via G. Gronchi 18, University of L'Aquila, I-67100 L'Aquila, Italy}
\author{Marcello Alecci}
\affiliation{Department of Life, Health and Environmental Sciences, University of L'Aquila, Via Vetoio 1, I-67100 L'Aquila, Italy} 
\affiliation{Institute for Superconductors, Oxides and Other Innovative Materials and Devices, National Research Council, Via Vetoio 1, I-67100 L'Aquila, Italy}
\affiliation{National Institute for Nuclear Physics (INFN), Gran Sasso National Laboratory, I-67100,  L'Aquila, Italy}

\begin{abstract}
A homogeneous negative permeability sphere can support magnetic localized surface plasmons (MLSPs). Generally, negative permeability materials are metamaterial (MM) structures exhibiting very deep subwavelength spatial scales, whose effects may be detrimental in the near-field for those applications based on effective medium approximations. We suggest to overcome this fundamental limitation by demonstrating analytically that the electromagnetic spatial distribution, associated to a MLSP resonance and excited by a near-field source, can be accurately reproduced outside the sphere by substituting the negative permeability sphere with a homogeneous high-index dielectric one with the same radius. Considering that a large class of ferroelectric materials shows ultra-high dielectric constant and low-losses at low frequency (up to GHz), our spoof MLSPs theory could be a key tool for realizing high performance subwavelength magnetic photonic devices in the radiofrequency and microwave regions. 
\end{abstract}

\maketitle
\section{Introduction}
Surface plasmon polaritons and localized surface plasmons are special wave modes tightly confined at the interface between two materials with opposite signs in the real part of the electric permittivities. They provide a unique platform for radiation manipulation on subwavelength scales \cite{Zaya}. At high frequencies (i.e., UV, visible, near-infrared), a metal can behave as a negative dielectric (i.e., the real part of dielectric constant is negative) and it can support surface plasmons excitation on its surface. Nevertheless, at lower frequencies (i.e., terahertz, microwave, and radio-frequency), metals do not support surface plasmons waves, since they behave almost as perfect electric conductors (where the permittivity is dominated by its imaginary part). To overcome this issue, Pendry \textit{et al.} \cite{Pendry_1} proposed metamaterial (MM) structures supporting the so-called spoof or designer surface plasmons. Indeed, they suggested that a MM with an effective negative permittivity can spoof the standard optical surface plasmons at the desired low frequency \cite{Yu,Liao,Qin,Chen,Huidobro_00,Li,Tang,Liao2,HuHu}. In addition, a suitable designed MM structure can exhibit artificial magnetic properties, thus an effective negative permeability MM can support spoof magnetic surface plasmon polaritons or spoof magnetic localized surface plasmons (MLSPs) \cite{Gollub,Liu,Hui,Liu_00,Gao,Liao_00,Rizza2}, the magnetic counterparts of the electric surface plasmon waves.  

In standard MM theories, the effective electromagnetic parameters (such as effective permittivity and permeability) describe the spatial dynamics of the macroscopic (average) fields at a scale much greater than the MM inclusions' size \cite{Alu,Rizza0}. As any result based on an effective medium approach, the standard spoof plasmon theory fails at very deep subwavelength scales and one can not accurately design and optimize photonic devices operating at these scales. Devilez \textit{et al.} \cite{Devilez} proposed an alternative approach to the standard effective medium theories. In fact, revisiting the standard plane-wave scattering approach (i.e., the Mie theory), they showed that a dielectric sphere (with positive permittivity), around its Mie resonances \cite{Mie1}, behaves effectively as a negative dielectric and, hence it can support spoof (electric) localized surface plasmons. Recently, we extended this result showing that a dielectric sphere with positive permittivity, around its Mie resonances, can support spoof MLSPs as well \cite{Rizza}. We investigated spoof MLSPs in the context of magnetic resonance imaging applications aiming to signal-to-noise ratio enhancement \cite{Rizza}.

In this paper, we demonstrate analytically that the desired MLSP electromagnetic spatial distribution, excited by means of a current carrying ring surrounding the magnetic permeability sphere, can be accurately reproduced by replacing the negative permeability sphere with a homogeneous high-index dielectric one with the same radius. The proof is based on exact solutions of Maxwell's equations, thereby our results hold in the extremely near-field region too.  Considering that high-index dielectric materials are available in nature in several frequency ranges, our analytical results open the way to realize spoof MLSPs-based devices without MM structuring, thus avoiding the fundamental limitations due to the effective medium approximations.  

The paper is organized as follows. In section 2, we analytically investigate the MLSPs resonances generated by a current carrying ring surrounding a negative permeability MM sphere. In section 3, we discuss the mimicking of the MLSPs resonances by considering the equivalent setup where a high-index dielectric sphere replaces the magnetic MM one. In section 4, we draw our conclusions.

\begin{figure}
\centering
\includegraphics[width=0.6\textwidth]{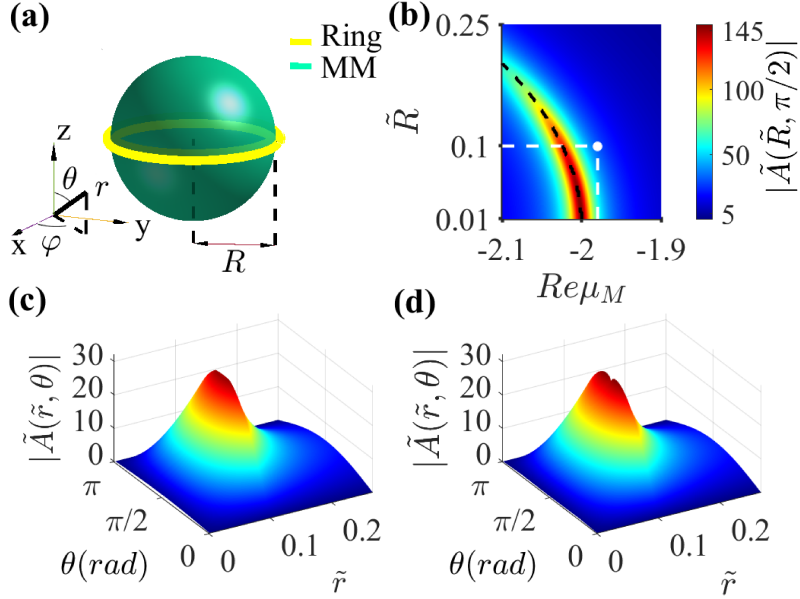}
\caption{(\textbf{a}) Sketch of the MLSPs setup. A negative permeability MM sphere with radius $R$ surrounded by a current carrying ring with same radius lying in the plane $z=0$. (\textbf{b}) $|\tilde A(\tilde R, \pi/2)|$ as a function of $\tilde R= k_0 R$  and $Re \mu_M$ evaluated for $Im \mu_M=0.01$ ($\epsilon_M=1$). The black dashed line corresponds to the $l=1$ MLSP resonance condition. The spatial profile of $|\tilde A|$ near the $l=1$ MLSP resonance condition with $\tilde R= 0.1$, $\mu_M=-1.98+i0.01$ (white point in panel b) analytically evaluated (\textbf{c}) by  neglecting the high multipolar contributions $l \ge 2$ [see Eq.(\ref{sol_par})] and (\textbf{d}) by considering full-wave simulations \cite{comsol,note1}, respectively.}
\label{Figure1}
\end{figure}
\section{Resonant excitations of MLSPs with a near-field source}
In the following, we consider MLSPs supported by a negative permeability MM with an ideal homogeneous isotropic electromagnetic response described by the dielectric and magnetic constants $\epsilon_M$ and $\mu_M$, respectively. The local field source is a current carrying ring with the same radius of the sphere $R$ and lying in the $z=0$ plane as shown in Fig.1(a). Exploiting both the MM spherical symmetry and the $z$-axis rotational invariance of the considered setup, we search for monochromatic solutions of the form 
$\vec{{  A}}=Re\left[ A(r,\theta) e^{-i \omega t }  \hat{\bf \varphi} \right]$,
with angular frequency $\omega$ and where $A(r,\theta)$ is the azimuthal component of the electromagnetic vector potential. Combining the Maxwell equations, we get the equation governing the field spatial dynamics, namely 
\begin{equation}
\label{helm}
\nabla^2 {\bf A}_{\varphi}+ \mu^{-1} \nabla \mu \times \left( \nabla \times {\bf A}_{\varphi} \right)  +  k_0^2 \epsilon \mu  {\bf A}_{\varphi}= - \mu_0 \mu {\bf J}_{\varphi}, 
\end{equation}     
where $\epsilon$ and $\mu$ are the relative dielectric permittivity and magnetic permeability, respectively, and ${\bf J}_{\varphi}$ is the ring density current ($k_0=\omega/c$, $c$ is the radiation vacuum speed and $\mu_0$ is the vacuum permeability). We suppose that the ring has negligible sizes along the radial and $z$ directions and thus it supports a current density ${J}_{\varphi}(r,\theta)=(I/R) \delta(\cos{\theta}) \delta(r-R)$, where $I$ is the ring current amplitude and $\delta(\cdot)$ is the Dirac delta function. By variables separation, we express the solutions of Eq.(\ref{helm}) in terms of spherical Bessel functions and associated Legendre polynomials, 
\begin{eqnarray}
\label{22}
 A && = \mu_0 I \sum_{l=1}^{+\infty}   P_l^{(1)}(\cos{\theta}) \cdot
 \begin{cases} 
   a_l j_l(\sqrt{\epsilon_M \mu_M} k_0 r)  \quad for \quad { r} \le { R}, \\
   
    b_l h_l^{(+)}(k_0 r)   \quad  for \quad { r} > { R},
\end{cases}
\end{eqnarray}
where $a_l$ and $b_l$ are series coefficients, $P_l^{(1)}$ is the associated Legendre polynomial $P^{(m)}_l$ with $m=1$, $j_l$ and $h_l^{(+)}$ are the spherical Bessel and the outgoing spherical Hankel functions, respectively. To evaluate the series coefficients $a_l$ and $b_l$, we use the expansion
\begin{equation}
\delta(\cos{\theta})=\sum_{l=1}^{+\infty} c_{l} \sqrt{\frac{2l+1}{2(l+1)l}} P_{l}^{(1)}(\cos{\theta}),
\end{equation}
where $c_{2l}=0$ and $c_{2l+1}=  \sqrt{\frac{4l+3}{4(l+1)(2l+1)}}P_{2l+1}^{(1)}(0)$ ($l=0,1,2,...$). Note that this expansion is derived from the completeness relation for the spherical harmonics \cite{Jackson} after multiplication by $e^{-i \varphi}$ and integration over $\varphi$. Imposing the interface conditions at the sphere surface $r=R$ for each $l$-mode [i.e., the continuity of the vector potential and, due to the ring current, the discontinuity of the polar magnetic field component], after some straightforward algebra, we get the $a_l$ and $b_l$ coefficients. As a consequence, the vector potential of Eq.(\ref{22}) can be written in the normalized form
\begin{eqnarray}
\label{sol_par}
 {\tilde A} && =  \sum_{l=1}^{+\infty}  \frac{c_l}{D_l(\epsilon_M, \mu_M, {\tilde R})}  \sqrt{\frac{2l+1}{2(l+1)l}} P_l^{(1)}(\cos{\theta}) \times \nonumber \\
&& \times \begin{cases}  
   j_l (\sqrt{\epsilon_M \mu_M} {\tilde r})/ j_l (\sqrt{\epsilon_M \mu_M} {\tilde R})  \quad for \quad {\tilde r} \le {\tilde R}, \\
   h_l^{(+)} ({\tilde r}) / h_l^{(+)} ({\tilde R})  \quad  for \quad {\tilde r} > {\tilde R},
   \end{cases} 
\end{eqnarray}
where 
\begin{equation}
\label{DDD}
D_l=\frac{1}{\mu_M} \phi_l^{(1)}(\sqrt{\epsilon_M \mu_M} {\tilde R})  -\phi_l^{(+)}({\tilde R}),  
\end{equation}
${\tilde A}({\tilde r},\theta)=A/( \mu_0 I)$, $\tilde r= k_0 r$, $\tilde R= k_0 R$, $\phi_l^{(+)}(\xi)=(d[\xi h_l^{(+)}(\xi)]/d \xi) / h_l^{(+)}(\xi)$, and $\phi_l^{(1)}(\xi)=(d[\xi j_l(\xi)]/d \xi) /j_l(\xi)$. One can demonstrate that the existence condition of a specific surface mode supported by the MM structure, without source terms, is $D_l=0$ for a given value of $l$. The equation $D_l=0$ shows exact solutions in the static limit (i.e. $\tilde R=0$) and it is satisfied for  
\begin{equation}
\label{existence_stat}
\mu_M=-\frac{1+l}{l}. 
\end{equation}
The equation (\ref{existence_stat}) represents the existence condition of the MLSPs for a lossless material in the static limit without source terms. In addition, it is worth noting that a MLSP excitation shows a resonant nature, since the relation $D_l=0$ yields a singularity in the series expansion of Eq.(\ref{sol_par}).  As an example, we consider an ideal homogeneous MM sphere with negative permeability and $\epsilon_M=1$. In Fig.1(b), we report $|\tilde A({\tilde R}, \pi/2)|$ as a function of $\tilde R$ and $Re \mu_M$ with $Im \mu_M=0.01$. As expected, near the value $Re \mu_M=-2$ (see Eq.(\ref{existence_stat})), we observe a resonance behavior associated to the $l=1$ MLSP excitation. In Fig.1 (c), we report the absolute value of the vector potential $|\tilde A(\tilde r, \theta)|$  of the resonant $l=1$ mode that has been evaluated analytically from Eq.(\ref{sol_par}) neglecting the higher multipolar order contributions $l \ge 2$. In panel (d), we plot $|\tilde A(\tilde r, \theta)|$ evaluated numerically by full-wave simulations with $\tilde R=0.1$ (more precisely, as an example, we set $R=3.0$ cm and thus the working frequency is $\nu=159.2$ MHz) \cite{comsol,note1}. Even though, as shown in Fig.1(b), the chosen permeability value $\mu_M=-1.98+i0.01$ is not optimized to achieve the MLSP resonance for $\tilde R= 0.1$, it is evident that the two spatial profiles are very similar as shown in Fig. 1(c) and 1(d). As a consequence, we conclude that the higher multipolar contributions [i.e. the $l\ge2$ terms in Eq.({\ref{sol_par}})] are very small. In other words, the $l=1$ MLSP resonance condition filters the other multipolar contributions and the electromagnetic field spatial distribution supported by the considered setup is very close to the spatial distribution of the resonant $l=1$ mode only. 

\section{Spoof MLSPs supported by a high-index dielectric sphere}
Here, we analytically investigate MLSPs mimicking. We search setups supporting surface waves with associated field distributions similar to MLSPs. Following the previous work \cite{Rizza}, we consider the configuration where the negative permeability MM sphere ($\epsilon_M=1$) is substituted by an equivalent one, with the same radius $R$, whose electromagnetic response is described by a permittivity $\epsilon_{eq}$ and a permeability $\mu_{eq}$. Generally, negative permeability materials are not available in nature at the desired frequency, so that mimicking is of particular interest when it can be realized by a sphere of a radius $R$ with $\mu_{eq}=1$ and $\epsilon_{eq}>0$, albeit our approach is more general. For a sphere of such material, the spatial distribution of the electromagnetic vector potential can still be written as in Eq.(\ref{sol_par}) (where the parameters with the subscript $'M'$ are substituted with the one having subscript $'eq'$). Our working hypothesis is that when both setups are near the same $l^*$-mode resonance (i.e., $D_{l^*} \simeq 0$ for both configurations), the other multipolar contributions ($l \neq l^*$) in the series expansion of Eq.(\ref{sol_par}) provide a negligible contribution. As a consequence, it is evident from Eq.(\ref{sol_par}) that the two considered $l^*$ modes (one hosted by the MM and the other one by the equivalent material) share the same electromagnetic field profile outside the sphere. Comparing the analytical solution of the two $l^*$-modes (namely, neglecting the $l \neq l^*$ modes and equaling the second of Eq.(\ref{sol_par}) of the two configurations), we obtain that the mimicking effect is given by the condition  
\begin{equation}
\label{existence_eq1}
\mu_{eq} \phi_{l^*}^{(1)}(\sqrt{\epsilon_M \mu_M} {\tilde R})-\mu_M \phi_{l^*}^{(1)}(\sqrt{\epsilon_{eq} \mu_{eq}} {\tilde R})=0.
\end{equation}
By solving numerically Eq.(\ref{existence_eq1}), we determine the equivalent electromagnetic parameters $\epsilon_{eq}$ and $\mu_{eq}$. 
\begin{figure}
\centering
\includegraphics[width=0.6\textwidth]{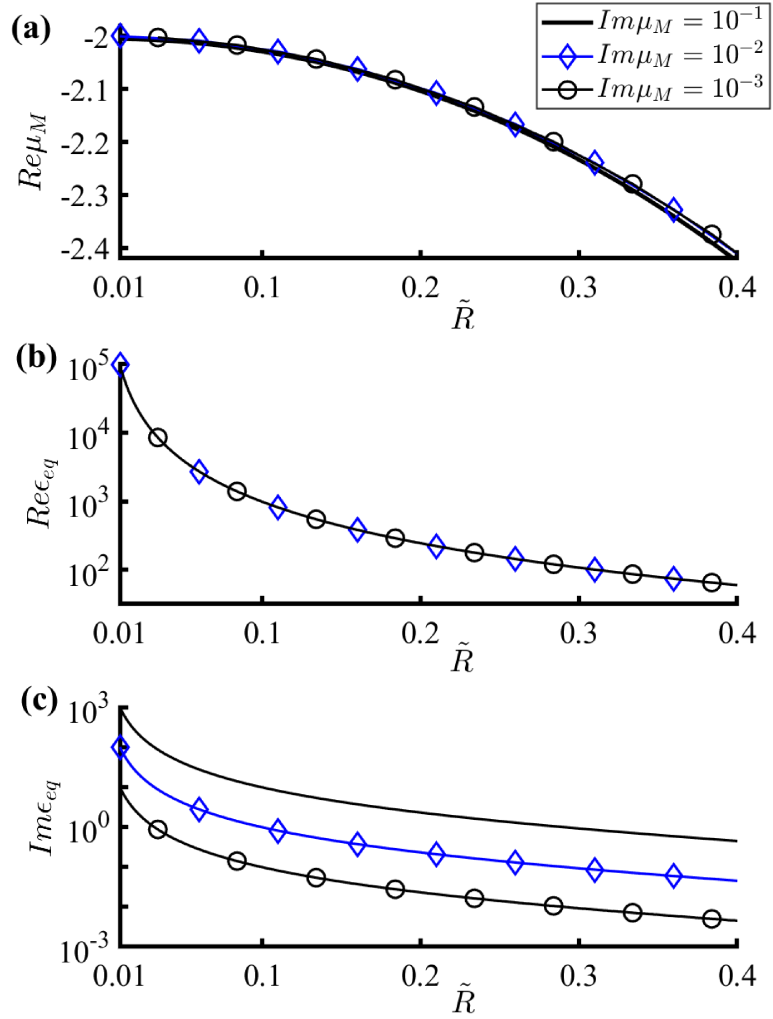}
\caption{(\textbf{a}) The real part of $\mu_{M}$ 
supporting the $l^*=1$ MLSP resonance (with $\epsilon_M=1$) as a function of $\tilde R$ for $Im \mu_M=10^{-1}$,$10^{-2}$,$10^{-3}$. The (\textbf{b}) real and the (\textbf{c}) imaginary part of $\epsilon_{eq}$ (with $\mu_{eq}=1$) equivalent to the $\mu_{M}$ reported in panel (a).} 
\label{Figure2}
\end{figure}
In the following, we focus our attention on the $l^*=1$ resonance, even though our results hold for higher $l^*$ values as well. When material losses are present, even in the quasi static regime (i.e., $\tilde R \ll 1$), $D_{l^*}$ can not exactly be zero, thus we define the $l^*=1$ resonance condition when the function $|D_1|$ approaches its minimum. 
In Fig.2 (a), we plot the value of $Re \mu_M$ (with $\epsilon_M=1$) supporting the $l^*=1$ MLSP resonance for $Im \mu_{M}=10^{-1},10^{-2},10^{-3}$ and $\mu_{eq}=1$. The corresponding mimicking dielectric constants $Re \epsilon_{eq}$, $Im \epsilon_{eq}$ (with $\mu_{eq}=1$) from Eq.(\ref{existence_eq1}) are plotted in Fig. 2(b) and (c), respectively. As shown in Fig.2(a) and (b), in the quasi-static regime, large values of equivalent dielectric $\epsilon_{eq}$ are necessary to mimic a negative permeability MM (e.g., for $\tilde R=0.01$, the sphere with $\epsilon_{eq} \simeq 10^5+i 10^3$ reproduces the MLSP resonance with $\mu_M \simeq -2.0 +i 0.1$). If we consider increasing $\tilde R$ values mimicking is realized by
decreasing dielectric constant materials (e.g., for $\tilde R=0.4$, $\epsilon_{eq} \simeq 60.0+i0.4$ is equivalent to $\mu_M \simeq -2.4 +i 0.1$). 

\begin{figure}
\centering
\includegraphics[width=0.6\textwidth]{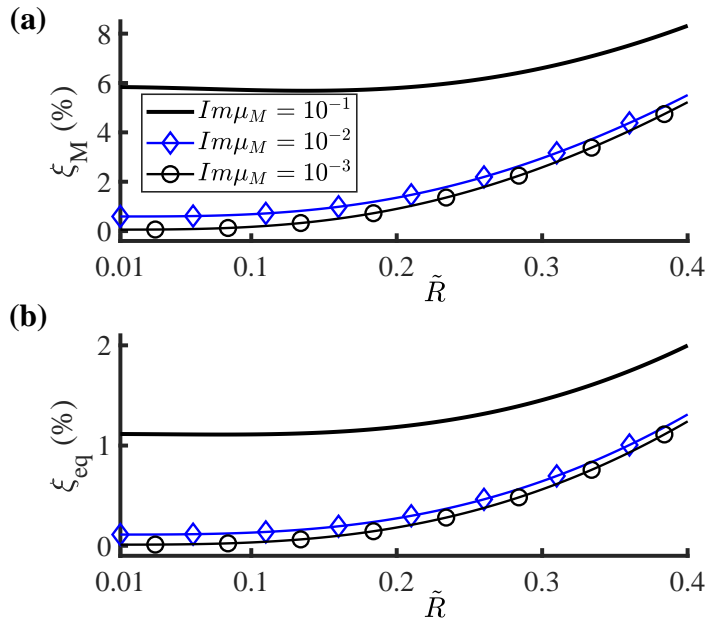}
\caption{(\textbf{a}) $\xi_M$ and (\textbf{b}) $\xi_{eq}$ quantify the deviation between high order multipolar terms ($l\ge 2$) with respect to the the magnetic dipolar one ($l^*=1$) for the MM and the equivalent dielectric material, whose electromagnetic parameters as in Fig.2.} 
\label{Figure3}
\end{figure}

To quantify the mimicking accuracy, we define 
\begin{equation}
\label{xiii}
\xi (\epsilon,\mu,{\tilde R}) =\sqrt{\sum_{l=2}^{L} \left| \frac{ c_{l} }{ D_{l} (\epsilon, \mu, {\tilde R}) }  \right|^2}.
\end{equation}
In Fig.3, we plot $\xi_M=\xi(\epsilon_M,\mu_M,{\tilde R})$ and $\xi_{eq}=\xi(\epsilon_{eq},\mu_{eq},{\tilde R})$, with $L=81$, expressed as percentage of $|c_1/D_1(\epsilon_M,\mu_M,{\tilde R})|$, $|c_1/D_1(\epsilon_{eq},\mu_{eq},{\tilde R})|$, respectively, for the electromagnetic parameters used in Fig.2. The results reported in Fig.3 provide the evidence that the mimicking approach works very well, within few percents, 
for small $\tilde R$ values and is less accurate for increasing $\tilde R$ values. In the considered $\tilde R$ range, the largest deviations are $\xi_M \simeq 8$ $\%$ and $\xi_{eq} \simeq 2$ $\%$ for the magnetic and the dielectric sphere, respectively. These results suggest that the $l=1$ MLSP mimicking can be realized in frequency regions where natural low-losses high-index materials ($Re\epsilon_{eq} \geq 60$, $Im \epsilon_{eq} \ll Re \epsilon_{eq}$) are available.  

\begin{figure}[h]
\centering
\includegraphics[width=0.6\textwidth]{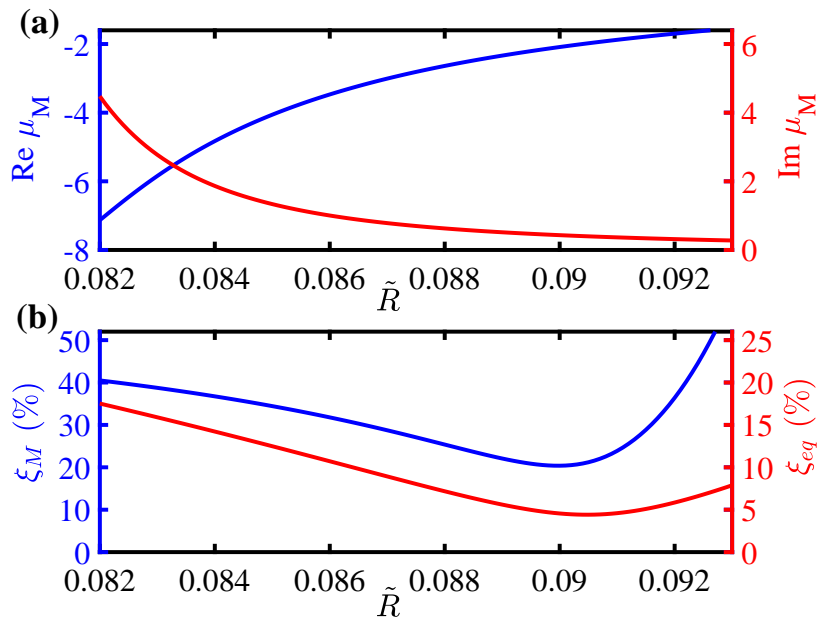}
\caption{(\textbf{a}) The $\mu_{M}$ values (with $\epsilon_M=1$) supporting the $l^*=1$ MLSP resonance at the frequency $\nu=127.7$ MHz corresponding to the equivalent electromagnetic case with $\epsilon_{eq}=1200+i 48$ and $\mu_{eq}=1$. (\textbf{b}) Percentage $\xi_M$ for the magnetic MM [$\mu_M$ and $\epsilon_M$ as in panel (a)] and $\xi_{eq}$ for the equivalent dielectric material ($\epsilon_{eq}=1200+i 48$, $\mu_{eq}=1$).} 
\label{Figure3}
\end{figure}

It is interesting to note that homogeneous low-loss high-index dielectric materials are available in nature in different spectral regions. They can support Mie resonances offering novel routes to manipulate electromagnetic radiation \cite{Staude,Tzarouchis,Mie0,Mie2,Mie3}. At low frequencies (up to GHz), ferroelectric materials show both a very-high real part of the permittivity and very-low electromagnetic losses \cite{Gevorgian,Webb_mater}. They represent natural candidates for supporting the proposed spoof MLSPs. For practical applications, it is interesting to reverse our approach, viz., starting from available high-index dielectric materials, we can determine their effective negative permeability.

Here, as a leading example, we consider a dielectric sphere with $\epsilon_{eq}=1200+i 48$ ($\mu_{eq}=1$) at the frequency $\nu=127.7$ MHz, corresponding to the dielectric constant of ceramics based on lead zirconate titanate (PZT), recently proposed for nuclear magnetic resonance imaging applications \cite{Rupprecht}. In Fig.4 (a), as a function of the dimensionless radius $\tilde R$, we report the $\mu_{M}$ values supporting the $l^*=1$ MLSP resonance, evaluated by using Eq.(\ref{existence_eq1}). In Fig.4 (b), we report the percentage $\xi_M$ (for the $\mu_M$ values reported in Fig.4 (a) and $\epsilon_M=1$) and $\xi_{eq}$ (for the above PZT dielectric values) as a function of dimensionless radius $\tilde R$.  As expected, in Fig.4(b), both $\xi_M$ and $\xi_{eq}$ show a minimum at about $\tilde R =0.09$, when $Re \mu_M$ is near $-2$, i.e. close to the $l^*=1$ resonant condition [see Eq. (\ref{existence_stat})].  In addition, we note that the percentage $\xi_{eq}$ is smaller than $\xi_{M}$, in other words, accordingly to the results showed in Fig.3, Fig.4(b) shows that the dielectric resonance is more effective than the magnetic one.
\begin{figure}[h]
\centering
\includegraphics[width=0.6\textwidth]{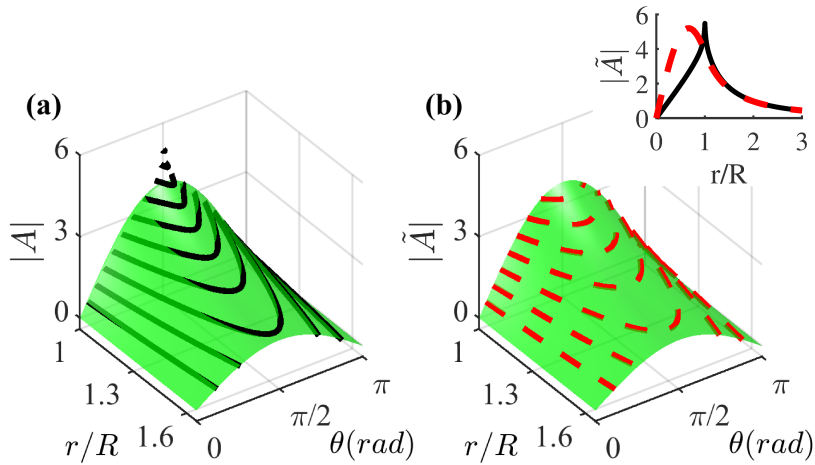}
\caption{The contour lines of $|\tilde A|$ outside the sphere \textbf{(a)} (solid black lines) with the magnetic MM sphere ($\mu_M=-1.98+i0.40$, $\epsilon_M=1$) and \textbf{(b)} (dashed red lines) with the equivalent dielectric one ($\epsilon_{eq}=1200+i 48$, $\mu_{eq}=1$). The vector potential is evaluated by full-wave simulations \cite{comsol,note1} for $\nu=127.7$ MHz, $R=3.38$ cm (i.e., $\tilde R = 0.0905$). For comparison purposes, we plot the corresponding $|\tilde A|$ (the green surfaces) of the $l^*=1$ MLSP supported by the magnetic MM, i.e. analytically evaluated by Eq.(\ref{sol_par}) neglecting the terms $l\ge 2$. In the inset, the $|\tilde A|$ profiles for $\theta=\pi/2$, for the magnetic (solid black line) and the equivalent dielectric configuration (dashed red line), are compared inside and outside the sphere.}
\label{Figure1}
\end{figure}
In Fig.5, we report numerical results for $\nu=127.7$ MHz, $R=3.38$ cm, i.e., $\tilde R = 0.0905$ \cite{comsol,note1}. We plot $|\tilde A|$ corresponding to the $l^*=1$ MLSP resonance ($\nu=127.7$ MHz, $R=3.38$ cm, ${\tilde R}=0.0905$) of the (a) negative permeability MM sphere ($\mu_M=-1.98+i0.40$, $\epsilon_M=1$) and (b) equivalent dielectric one ($\epsilon_{eq}=1200+i 48$, $\mu_{eq}=1$), respectively, as a function of the polar angle $\theta$ and the ratio between the radial position (outside the sphere) and the sphere radius (i.e., $r/R \ge 1$). The vector potential in both configurations, evaluated by full-wave simulations, is compared with the one evaluated by Eq.(\ref{sol_par}) neglecting the $l \ge 2$ terms. For a more qualitative comparison, in the Fig.5 inset, we report the $|\tilde A|$ profile with $\theta=\pi/2$, for the magnetic and the equivalent dielectric sphere, as a function of $r/R$. The MLSPs mimicking shows a fairly good accuracy, since the deviation is tightly localized near the ring location (i.e., $\tilde r=\tilde R= 0.0905$ and $\theta=\pi/2$). In fact, at $\tilde r= 0.0905$ and $\theta=\pi/2$, the absolute value of the vector potential is $5.5$, $4.0$, $3.8$ for the magnetic sphere, the dielectric one and the $l^*=1$ mode, respectively. In agreement with the results of Fig.4(b), the dielectric sphere reproduces the $l^*=1$ mode better than the negative permeability MM one.   

\section{Conclusions}
In summary, we investigate analytically MLSPs resonances supported by a negative permeability MM sphere and excited by a current carrying ring with the same radius. We prove that the considered setup, where the MM sphere is substituted by a homogeneous high index dielectric one, is able to support an accurate MLSP mimicking outside the sphere at extremely near-field scales. We stress that our theoretical model hold also for high-order MLSPs and it could be generalized to other geometries for achieving the desired effective magnetic permeability. Considering the current research efforts on novel high performance high-index dielectric materials at the radio-frequencies and microwaves spectral regions, our results paves the way to the design of subwavelength magnetic photonic components useful for several applications, such as nuclear magnetic resonance spectroscopy, imaging, biosensing.

\section{Acknowledgments} 
This work has been supported by the CNR-SPIN Nano-Agents project.


\begin{thebibliography}{99}
\bibitem{note1} Exploiting the setup $z$-axis rotational invariance, we performed 2D full-wave simulations in cylindrical coordinates (i.e. in the $(\rho,z)$ plane, where $\rho=r \sin{\theta}$, $z=r\cos{\theta}$). The current carrying ring was modeled with a surface current density at the sphere surface $J_{\varphi}={K}_{\varphi} \delta(r-R)$, where ${K}_{\varphi}=(I/R)exp{\left(-\frac{z^2}{R^2 \sigma^2}\right)}/{\sqrt{\pi \sigma^2}}$ and $\sigma=10^{-2}$. The considered structure was surrounded by vacuum and a simulation domain region with open boundaries was used (scattering boundary conditions, spherical waves type).
\end{thebibliography}
\end{document}